\documentclass[8pt, twocolumn]{extarticle}

\usepackage{amssymb,amsfonts,amsmath}

\usepackage{booktabs}
\usepackage{graphicx}
\usepackage{authblk}
\usepackage{geometry}
\usepackage{balance}
\usepackage{pdfpages}

\usepackage{lettrine}

\usepackage{flushend}
\usepackage{bm}
\usepackage{color}

\newcommand{\rboxed}[1]{\fbox{$\displaystyle#1$}}
\makeatother
\renewcommand{\rboxed}[1]{\textcolor{red}{%
	\fbox{\normalcolor$\displaystyle#1$}}}
	
\newcommand{\bboxed}[1]{\fbox{$\displaystyle#1$}}
\makeatother
\renewcommand{\bboxed}[1]{\textcolor{blue}{%
	\fbox{\normalcolor$\displaystyle#1$}}}

  \newcommand\blfootnote[1]{%
  \begingroup
  \renewcommand\thefootnote{}\footnote{#1}%
  \addtocounter{footnote}{-1}%
  \endgroup
}

\definecolor{mPurp}{RGB}{126 , 47, 142}
\definecolor{mRed}{RGB}{217 , 83, 25}
\definecolor{mBlue}{RGB}{0 , 114, 189}
\definecolor{mGreen}{RGB}{119 , 172, 48}

\AtBeginDocument{}

\begin{document}
\title{\huge{\textsf{\textbf{Intracycle Angular Velocity Control of Cross-Flow Turbines}}}}

\author{\textsf{\textbf{Benjamin Strom*, Steven L. Brunton, and Brian Polagye}}}
\affil{ \textsf{\textbf{\normalsize Dept. of Mechanical Engineering, University of Washington}}\\  \textsf{\textbf{\normalsize Box 352600, NE Stevens Way, Seattle, WA, 98195, USA}}}
\date{}
\twocolumn[
\begin{@twocolumnfalse}
\maketitle
	\begin{abstract}
\textbf{\textsf{
Cross-flow turbines, also known as vertical-axis turbines, have numerous features that make them attractive for wind and marine renewable energy.  To maximize power output, the turbine blade kinematics may be controlled during the course of the blade revolution, thus optimizing the unsteady fluid dynamic forces. Dynamically pitching the blades, similar to blade control in a helicopter, is an established method. However, this technique adds undesirable mechanical complexity to the turbine, increasing cost and reducing durability.  Here we introduce a novel alternative requiring no additional moving parts: we optimize the turbine rotation rate as a function of blade position resulting in motion (including changes in the effective angle of attack) that is precisely timed to exploit unsteady fluid effects.  We demonstrate experimentally that this approach results in a 79\% increase in power output over industry standard control methods. Analysis of the fluid forcing and blade kinematics show that maximal power is achieved through alignment of fluid force and rotation rate extrema.  In addition, the optimized controller excites a well-timed dynamic stall vortex, as is found in many examples of biological propulsion.  This control strategy allows a structurally robust turbine operating at relatively low angular velocity to achieve high efficiency and could enable a new generation of environmentally-benign turbines for wind and water current power generation.}}

	\end{abstract}
	\end{@twocolumnfalse}
	]

\renewcommand{\LettrineTextFont}{\normalfont}
	
 \lettrine[lines=2,lraise=0]{\textsf{T}}{} he prospects of energy security and climate change continue to drive development of renewable energy sources. 
Kinetic energy from wind and water is abundant, and the conversion of this energy to rotational mechanical energy and, subsequently to electrical energy, has tremendous potential to power our modern world.  
Wind turbines are among the most cost effective and fastest growing sources of renewable energy~\cite{WEO_2015}. 
During the birth of the modern wind energy industry, a variety of turbine configurations were considered before designs converged on the now ubiquitous axial-flow turbine, otherwise known as a horizontal-axis turbine. 
Vertical-axis wind turbines, in which a set of blades rotate around an axis perpendicular to the direction of the free stream flow, provide an alternative design~\cite{Eriksson_2008,Bhutta_2012}; more generally, this type of turbine is referred to as a cross-flow turbine, as the rotation axis may not be vertical in some marine applications. 
Research on hydrokinetic turbines, where the turbine operates underwater, has increased in recent years. Like the early years of the wind energy industry, the nascent marine and fluvial hydrokinetic energy industry is searching for optimal turbine designs~\cite{Khan_2009}.    
	
	\begin{figure*}[t]
\vspace{-.35in}
\begin{center}
\includegraphics[width=\textwidth]{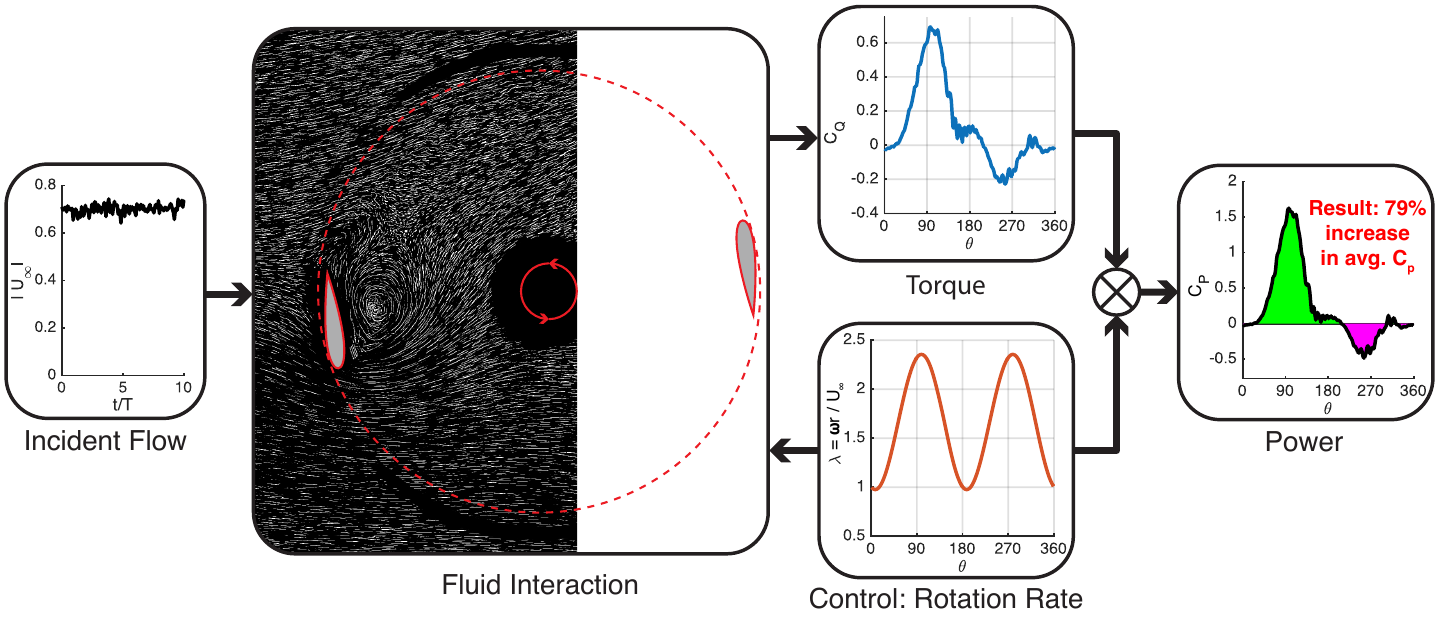}
\caption{\small \textsf{ An illustration of the interdependence of the control kinematics (angular velocity), the fluid structure interaction, forcing, and resulting power output. In the case where angular velocity is the kinematic control parameter, the timing of the angular velocity profile not only affects the fluid forcing by changing the local flow structure, but also directly affects the power output. An effective angular velocity controller then maximizes beneficial fluid structure interaction and aligns the highest angular velocity with the highest fluid torque. Note that here the input rotation rate $\omega$ is normalized by the free stream velocity $U_{\infty}$ and the radius $r$, resulting in the tip speed ratio $\lambda$.  The torque used to calculate the instantaneous torque coefficient and efficiency ($C_Q$ and $C_P$ respectively) does not include the torque necessary to accelerate and decelerate the turbine. Due to the periodic nature of the accelerations, these torques do not contribute to the mean power output (see Sup. Fig. S5). The areas lacking streaks in the flow visualization image are due to diffraction from the edge of the lower turbine end-plate. }}\label{fig:block}
\end{center}
\end{figure*}
\blfootnote{*email: strombw@uw.edu}
	A recent resurgence of research and commercial interest in cross-flow turbines is motivated by several factors. First, a vertically oriented cross-flow turbine operates omni-directionaly, removing the need for active yaw control. This is beneficial for urban wind generation, where the wind direction is often variable~\cite{Balduzzi_2012}, and in reversing tidal currents. Second, in a vertical orientation, heavy turbine components, such as the gearbox and generator, can be located at the base. For offshore wind applications, this increases the stability of the floating platform and ease of access for repair~\cite{Sutherland_2012}. Third, arrays of counter-rotating cross-flow turbines may be able to outperform equivalently sized arrays of axial-flow turbines due to beneficial device-device interaction~\cite{Whittlesey_2010,Dabiri_2011,Kinzel_2012}. Fourth, cross-flow turbines are more suited to exploit the energy of moving water in natural channels~\cite{Garrett_2007} since their rectangular projected area enables high-blockage configurations~\cite{Salter_2012}. Finally, cross-flow turbines generally operate at a lower relative rotational velocity than axial-flow turbines~\cite{Khan_2009}. 
These low relative velocities have socio-environmental benefits, including decreased noise and vibration, limited cavitation in hydrokinetic applications, and reduced risk of collision with avian and aquatic species.  
	However, reducing rotational velocity can reduce efficiency, providing an economic disincentive for this practice.
	
	In their most primitive form, cross-flow turbines have one degree of freedom: rotation about their central axis. Despite this apparent simplicity, the turbine blades encounter a broad range of conditions over the course of a single rotation, including large variations in the effective angle of attack experienced by the blade.
In some operating conditions, the flow separates over the blade, leading to the formation of leading-edge vorticity~\cite{Birch2001nature,videler:04lev,Muijres_2008,Chen:2010}.  
Drawing inspiration from biology, where flying and swimming animals achieve exceptional performance by harnessing similar unsteady flow structures, it is possible to enhance turbine performance by actively controlling the blade kinematics. Previous approaches have pitched the blade mechanically, much as a helicopter will pitch blades down during advance and up during retreat to achieve balanced loads.  
Here we apply a blade position based angular rotation rate controller that requires no additional degrees of freedom (i.e., no additional moving parts) and instead optimizes the blade's \emph{effective} angle of attack. Because mechanical power is the product of torque and angular velocity, this approach directly controls one of these variables and maximizes the power extraction during periods of largest fluid forcing. Figure \ref{fig:block} provides a high-level overview of this process. An added benefit 

This control strategy allows a structurally robust turbine that rotates at relatively low angular velocity to achieve a high efficiency. This could enable a new generation of environmentally-benign turbines for wind and water current power generation.

	The primary metric for cross-flow turbine performance is the conversion efficiency from the kinetic energy available in the projected area of the free stream flow to the mechanical power output of the turbine. This is given by the efficiency (also known as the power coefficient), 
	\begin{equation}
		C_P = \frac{\tau \omega}{\frac{1}{2} \rho U_\infty^3 A},
	\end{equation}
	where $\tau$ is the torque imparted to the turbine by the fluid, $\omega$ is the turbine rotation rate, $\rho$ is the density of the working fluid, $U_{\infty}$ is the free stream velocity, and $A$ is the projected area transverse to the free stream direction, which is the product of diameter and blade span for constant radius cross-flow turbines.

\begin{figure}[t]
\begin{center}
\includegraphics[width=0.9\columnwidth]{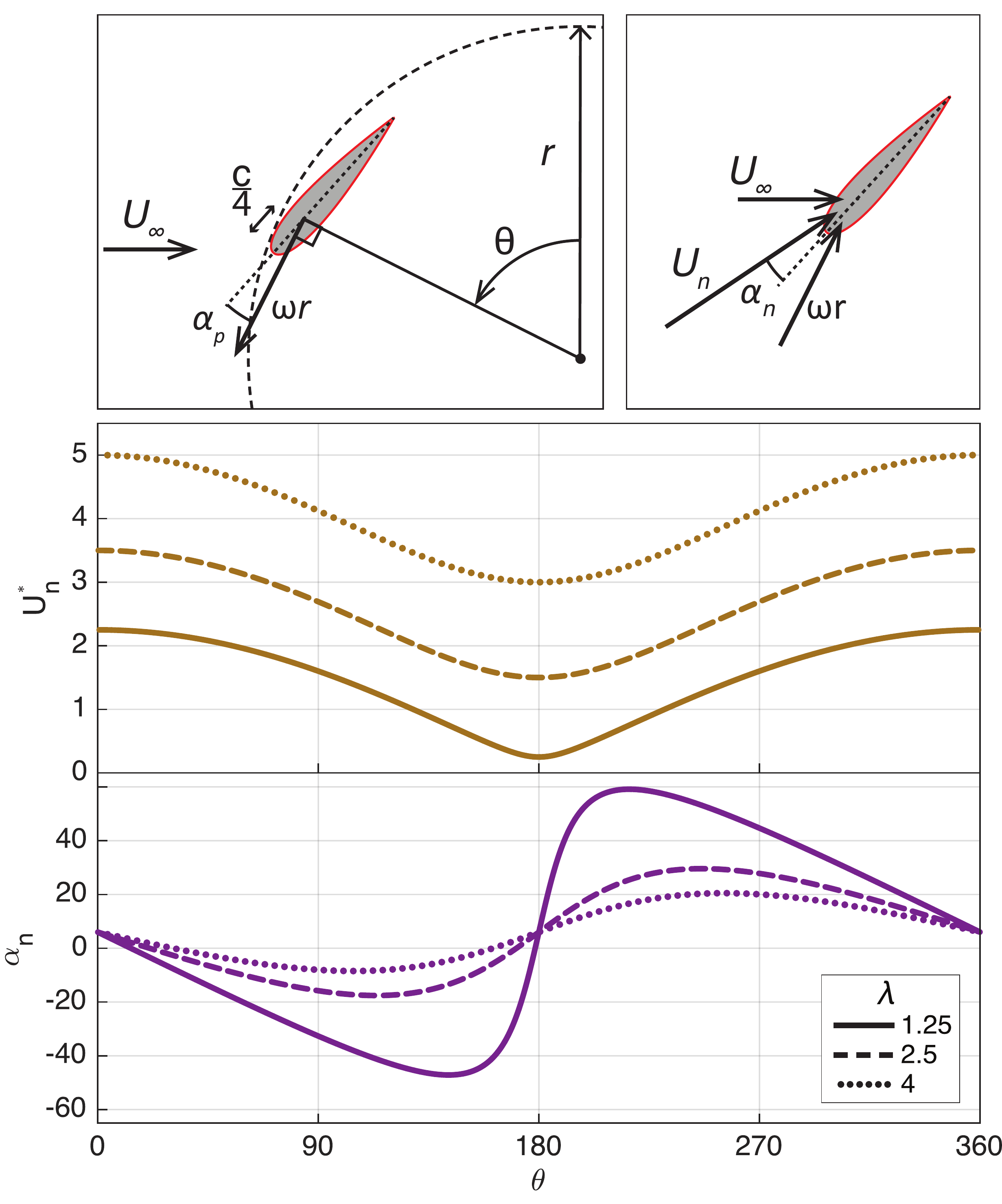}
\caption{\small \textsf{ Top left, schematic definition of the blade pitch angle, $\alpha_p$, as measured at the quarter chord, $c/4$, the azimuthal blade position, $\theta$ and the rotational velocity vector, $\omega r = \frac{d \theta}{dt} r$. The zero position, $\theta=0$, is defined for the foil traveling directly upstream. The top right defines the vector sum of the free stream velocity and the angular velocity as the nominal (effective) velocity, $U_n$. Center plot: The nominal velocity profiles for a turbine operating at three constant tip speed ratios. Bottom: The nominal angle of attack profiles for the same three tip speed ratios. }}\label{fig:intro}
\end{center}
\end{figure}

	{Torque is generated by components of the lift and drag forces on the turbine blades that are aligned with the direction of rotation.}  
The lift and drag forces are likewise governed by the angle of attack, its rate of change, and the relative flow velocity in the reference frame of the foil~\cite{Brunton2014jfs}. 
These values depend on the free stream flow velocity, the angular velocity of the foil, and any velocities induced on the flow field by the turbine. If the induced velocities are neglected, the nominal, or \emph{effective}, angle of attack can be written as a function of the azimuthal blade position, $\theta$:
\begin{equation}\label{eq:alpha_n}
		\alpha_n(\theta) = \tan^{-1}\left(\frac{\sin(\theta)}{\rboxed{\lambda(\theta)} + \cos(\theta)}\right)-\bboxed{\alpha_p(\theta)}.
\end{equation}
Thus, the effective angle of attack depends only on the blade pitch angle $\alpha_p$ and the tip speed ratio $\lambda$:
		\begin{equation}\label{eq:lambda}
		\lambda(\theta) = \frac{\rboxed{\omega(\theta)}\;r}{U_\infty}.
\end{equation}
This is the ratio of the blade velocity to the free stream flow velocity, where $r$ is the turbine radius. In their most general form, $\alpha_p$ and $\omega$ can be functions of $\theta$, but in most cross-flow implementations both are held constant\footnote{In constant torque control, the rotation rate may vary slightly. Control set points may be altered to adapt to the free stream conditions, but these changes are slow compared to the turbine rotation rate.}.
The diagrams in Fig.~\ref{fig:intro} illustrate the relationships between these parameters\footnote{When the chord to radius ratio is relatively high, as is the case for this study, the nominal angle of attack at a given azimuthal position varies along the chord length. The effect may be studied via conformal mapping to a rectilinear flow field, which introduces a virtual camber to the blade~\cite{Migliore_1980}. The curvilinear flow field may also significantly impact the trajectories of coherent flow structures generated during dynamic stall and their contribution to forces on the blade~\cite{Tsai_2014}.}.

	With the same simplifications, the nominal magnitude of the relative velocity encountered by the foil can be written as
	\begin{equation}
		U_n(\theta)^* = \frac{| \bm{U_n}(\theta) |}{U_\infty} = \sqrt{\lambda(\theta)^2+2\lambda(\theta) \cos(\theta)+1}, \label{eq:U_n}
	\end{equation}
	where we have chosen the free stream velocity as a normalization factor. 
	
	As evidenced by Fig.~\ref{fig:intro}, when the tip speed ratio $\lambda$ and the blade pitch angle $\alpha_p$ are held constant, the foil experiences a virtual pitch up, pitch down maneuver.  Depending on the tip speed ratio, the range of nominal angles of attack experienced by the foil can far exceed the static stall angle at which flow separates from a stationary foil. However, during a rapid increase in angle of attack, the flow remains partially attached at larger angles than in the static foil case.  
	This phenomena is known as ``dynamic stall'' and is well studied in the context of helicopter blade aerodynamics~\cite{Leishman_2006}. The delay in flow separation is accompanied by an increase in the lift force (well above the maximum static lift value) followed by an increase in the drag force when separation finally occurs~\cite{Mccroskey_1981}. Dynamic stall is often associated with the formation of a leading edge vortex, caused by the roll up of a shear layer formed by the separated flow~\cite{Carr_1977}.  The interaction of the leading edge vortex with the blade may influence turbine performance. Dynamic stall  has been extensively demonstrated and studied in cross-

\noindent   flow turbines~\cite{Buchner_2015,Ferreira_2007,Fraunie_1986,Fujisawa_2001}. 
	While dynamic stall is sometimes considered undesirable for cross-flow turbine operation~\cite{Buchner_2015}, birds~\cite{Warrick_2005}, bats~\cite{Muijres_2008}, and insects~\cite{Dickinson_1999} have all been shown to execute flight maneuvers that exploit the dynamic stall process and resulting vortical structures. Dynamic stall has been used to maximize the objectives of engineering problems including the lift of a flapping flat plate~\cite{Milano_2005}, the thrust of an oscillating foil~\cite{Hover_2004}, and the power produced by a pitching and heaving foil~\cite{Kinsey_2008,Ashraf_2011,Xiao_2012,Strom_2014}. This suggests that cross-flow turbines may also be able to benefit from dynamic stall.

	Control based on the angular position of a cross-flow turbine blade (i.e., $\alpha_p(\theta)$, $\omega(\theta)$) is referred to here as intracycle control to differentiate it from schemes that optimize turbine power over longer time scales in response to changes in the free stream velocity, such as~\cite{Munteanu_2009,Johnson_2006}. Approaches to intracycle control of cross-flow turbines can be split into two categories: schemes that alter the turbine kinematics and those that apply flow control to the foil surface to eliminate or delay separation of the upper foil boundary layer at high angles of attack (e.g., plasma actuators~\cite{Greenblatt_2014}, synthetic jets~\cite{Yen_2013}). While flow control has demonstrated benefits, these types of actuators would be difficult and expensive to implement commercially~\cite{Brunton2015amr}. Here, we demonstrate a kinematic intracycle controller that exploits the benefits of dynamic stall (as in the enumerated biological inspirations) rather than attempting to suppress it. The primary kinematic intracycle control scheme studied to date is active pitch control, in which the blade pitch angle varies as a function of angular position. This alters the nominal angle of attack via \eqref{eq:alpha_n}, where $\alpha_p$ is a function of $\theta$.

	\begin{figure}[t]
\begin{center}
\includegraphics[width=0.85\columnwidth]{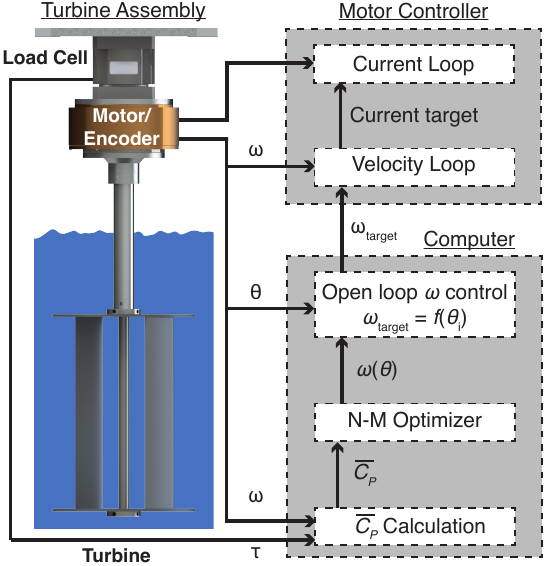}
\caption{\small \textsf{ Schematic of the experimental setup and diagram of the control scheme, as well as the Nelder-Mead optimization  procedure.}}\label{fig:setup}
\end{center}
\end{figure}

	This concept has been well studied both experimentally and numerically~\cite{Kirke_2011,Paraschivoiu_2009,Schonborn_2007}.  Increases in turbine efficiency of up to 24\% have been demonstrated in experiments~\cite{Kirke_2011}. However, these methods have not been commercially adopted, perhaps because the increase in mechanical complexity outweighs the efficiency gains.

\begin{figure*}[t]
\begin{center}
\noindent
\hspace{-0.15in}
\begin{minipage}{.7\textwidth}
\begin{tabular}{llccc}
\toprule
\small \textbf{Control Scheme} &\small \textbf{Control Parameters}
& \small  $ \boldsymbol{\overline{C_P}}$ & \small  $\boldsymbol{\sigma(C_P)}^{\text{\textbf{[a]}}}$ & \small \textbf{Increase}$^{\text{\textbf{[b]}}}$ \cr
\midrule[0.08em]
\textbf{\textcolor{mRed}{\small Constant $\boldsymbol{\tau}$ }}& \small $\tau =$ \small 0.093 N-m  & \textbf{\textcolor{mRed}{\small 0.219}} & \small 0.0057 & \cr \midrule
\textbf{\textcolor{mBlue}{\small Constant $\boldsymbol{\omega}$}  }& $ \omega =$ \small 15.46 rad/s & \textbf{\textcolor{mBlue}{\small 0.229}} & \small 0.0100 & \small \textit{Comparison}\cr \midrule
\textbf{\textcolor{mGreen}{\small Sinusoidal $\boldsymbol{\omega}$ }}& \small $\omega$ = 13.7 + 5.7 $\sin($2$\theta$ + 4.44$)$ rad/s & \textbf{\textcolor{mGreen}{\small 0.321}} & \small 0.0189 & \textbf{40\%} \cr \midrule
\textbf{\textcolor{mPurp}{\small Semiarbitrary $\boldsymbol{\omega}$}}  & \small $\omega$  = 15.8 + 6.9 $ \sin($2$\theta$ + 3.77$)$ +  & \textbf{\textcolor{mPurp}{\small 0.410}} & $\underline{\;\;\;\;\;}$$^{\text{\textbf{[c]}}}$ & \textbf{79\%} \cr
& \small $ \;\;\;\;\;\;\;$ 2.8 $\sin($4$\theta$ + 0.26$)$ +  & & & \cr
& \small $ \;\;\;\;\;\;\;$ 1.4 $\sin($6$\theta$ + 3.46$)$  rad/s & & & \cr
\bottomrule
\end{tabular}
{ \footnotesize
\begin{itemize}
	\item [\scriptsize{$^{\text{\textbf{[a]}}}$}] \hspace{-0.09in} Standard deviation of $C_P$ among turbine revolutions.
	\item [\scriptsize{$^{\text{\textbf{[b]}}}$}] \hspace{-0.09in} Percent increase in $\overline{C_P}$ in comparison to constant angular velocity control
	\item [\scriptsize{$^{\text{\textbf{[c]}}}$}] \hspace{-0.09in} Approximately uniform distribution between $C_P=0.33$ and $C_P = 0.50$.
\end{itemize}
}
\end{minipage}\hfill
\begin{minipage}{.25\textwidth}
\centering
\includegraphics[width=0.95\textwidth]{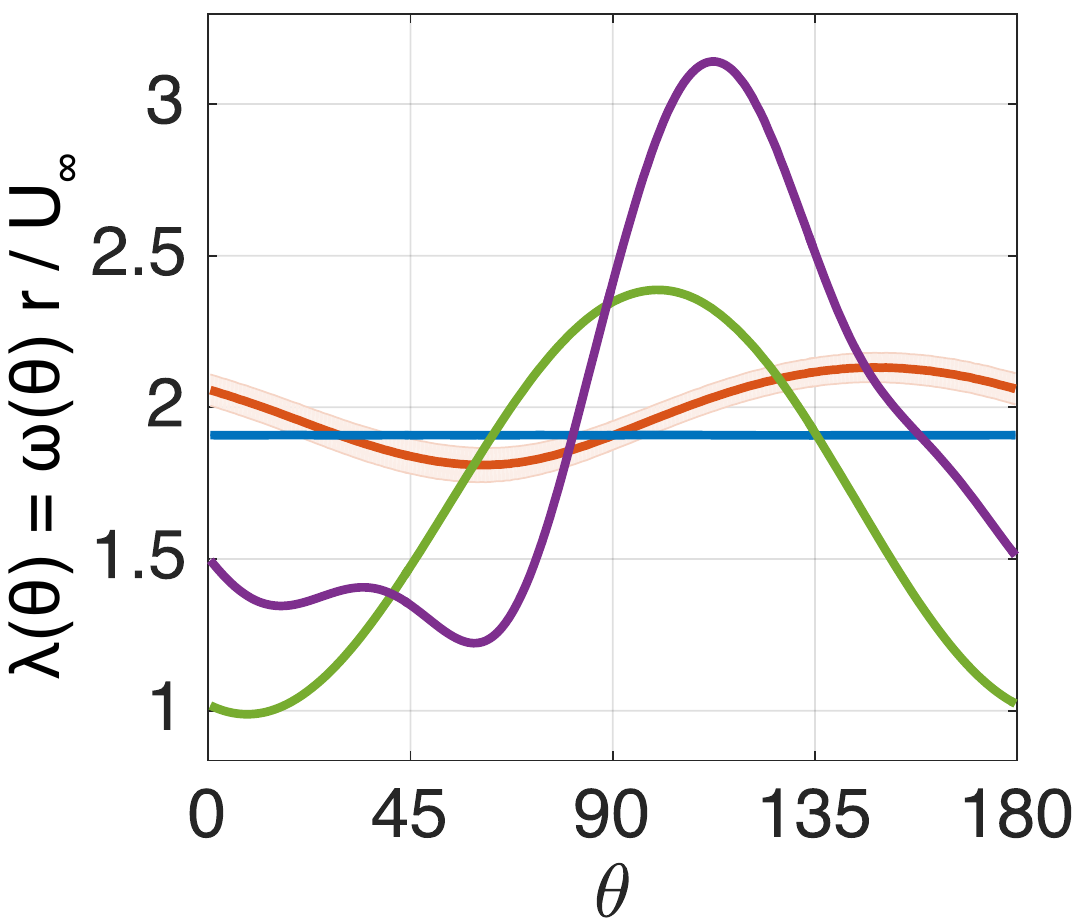}
\vspace{0.4in}
\end{minipage}
\vspace{-.1in}
\caption{\small \textsf{ Left: Optimum control parameters for the schemes tested, as well as their respective mean efficiencies ($\overline{C_P}$). The semiarbitrary control scheme shows a 79\% increase in efficiency over the constant angular velocity controller. Right: Tip speed ratio profiles of the optimum control schemes. A half revolution is presented as the profiles are twice periodic over a single turbine revolution. Note that the mean efficiency values presented are identical whether the total or fluid torque is used due to the angular velocity periodicity (see Sup. Fig. S5). }}\label{fig:results}
\end{center}
\end{figure*}

	As is evident from Eq.~\eqref{eq:alpha_n}, an alternative to a pitch control is to modify the turbine rotation rate (and thus the tip speed ratio) as a function of azimuthal blade position. This method may be implemented without increasing the mechanical complexity of the cross-flow turbine. Here, we explore the performance implications of two angular velocity profile parameterizations: a sinusoidal profile
	\begin{equation}
		\omega(\theta) =A_0 +A_1 \sin(N \theta + \phi_1),\label{Eq:Intracycle1}
	\end{equation}
	and a semi arbitrary profile (truncated Fourier series) 
	\begin{equation}
		\omega(\theta) = A_0 + \sum\limits_{i=1}^3 A_i \sin(i N \theta + \phi_i).\label{Eq:Intracycle2}
	\end{equation}
The frequency is a multiple of the number of blades $N$ to enforce periodicity and ensure that each blade experiences identical kinematics.  If the free stream velocity $U_\infty$ is quasi-steady, this is equivalent to controlling the tip speed ratio $\lambda(\theta)$.  The performance of these control schemes is compared to a turbine operating under two standard control methods: constant torque and constant angular velocity control.

\section{Controller Implementation}

	Turbine control tests were performed in a recirculating water flume with a test section measuring 75 cm wide and 47.5 cm deep. The resulting ratio of the test section area to the turbine cross-sectional area (known as the blockage ratio) was 11\%. The free stream velocity, continuously measured at a rate of 32 Hz by an acoustic Doppler velocimeter five diameters upstream from the turbine rotation axis, was maintained at 0.7 m/s. The incoming turbulence intensity was 2\%. The experimental Reynolds number, based on the chord length and free stream velocity, was 31,000. As shown in Fig.~\ref{fig:setup}, the cantilevered turbine was controlled using a servo motor that included a $10^6$ counts-per-revolution encoder. Reaction forces and torques were measured using a six-axis load cell.
	
	The turbine consisted of two straight NACA0018 foils, each with a chord length of 4 cm and a span of 23.4 cm, mounted to circular endplates at a pitch angle of 6$^\circ$ (leading edge rotated outward). The diameter of the turbine was 17.2 cm. The solidity, calculated as the fraction of the circumference occupied by blades, was 15\%. See Sup. Fig. S1 for a detailed turbine schematic.

	Force, torque, and angular position measurements  were taken at a sample rate of 1 kHz over a period of 30 seconds for a given turbine control parameter set. For constant torque and angular velocity control, control actuation that maximized $C_P$ was identified by incrementing the respective values (see Sup. Fig. S2 for the corresponding performance curves). Values for $A_i$ and $\phi_i$ that maximized $C_P$ under intracycle angular velocity control in Eqs.~\eqref{Eq:Intracycle1} and \eqref{Eq:Intracycle2} were selected by the Nelder-Mead (downhill simplex) method~\cite{Nelder_1965}.  This optimization was chosen due to the small number of required function evaluations and inspired by~\cite{Miller_2014}. The control and optimization procedure, shown in Fig.~\ref{fig:setup}, consisted of first evaluating the mean turbine performance at a given control parameter set, then incrementing the parameter set as required by the Nelder-Mead algorithm, and finally implementing the new parameter set. The optimizations were performed five times with randomized starting conditions to ensure that the solutions converged to a global maximum (see Sup. Fig. S3 plots of the optimization process). For comparison, the optimized versions of the four control schemes (constant torque, constant, sinusoidal, and semi-arbitrary angular velocity) were each tested ten times, alternating control schemes for each test, thus reducing the risk of the influence of changes in test conditions.  
	Intracycle angular velocity control requires angular accelerations and decelerations of the turbine. The corresponding torque fluctuations integrate to zero over one rotation due to the periodicity of the angular velocity profiles and are removed from instantaneous turbine performance quantities as:

\begin{equation}\label{eq:tfluid}
		\tau = \tau_{\text{fluid}} = \tau_{\text{total}} - I \ddot{\theta}
	\end{equation}
where $I$ is the rotational moment of inertia of the turbine. With this adjustment, the instantaneous fluid efficiency and torque coefficient may be calculated as a function of azimuthal blade position (see Sup. Fig. S4 and S5 for more details on the separation of the angular acceleration and the fluid torque) . 

	\section{Results}
	Optimized control scheme parameters, resulting tip speed ratio profiles, and their respective efficiencies are given in Fig.~\ref{fig:results}. Intracycle angular velocity control produced a substantial increase in turbine efficiency. Compared to the constant velocity control case, the sinusoidal and semi-arbitrary angular velocity control schemes yielded a 40\% and 79\% increase in efficiency, respectively.

\begin{figure}[t]

\begin{center}
\includegraphics[width=\columnwidth]{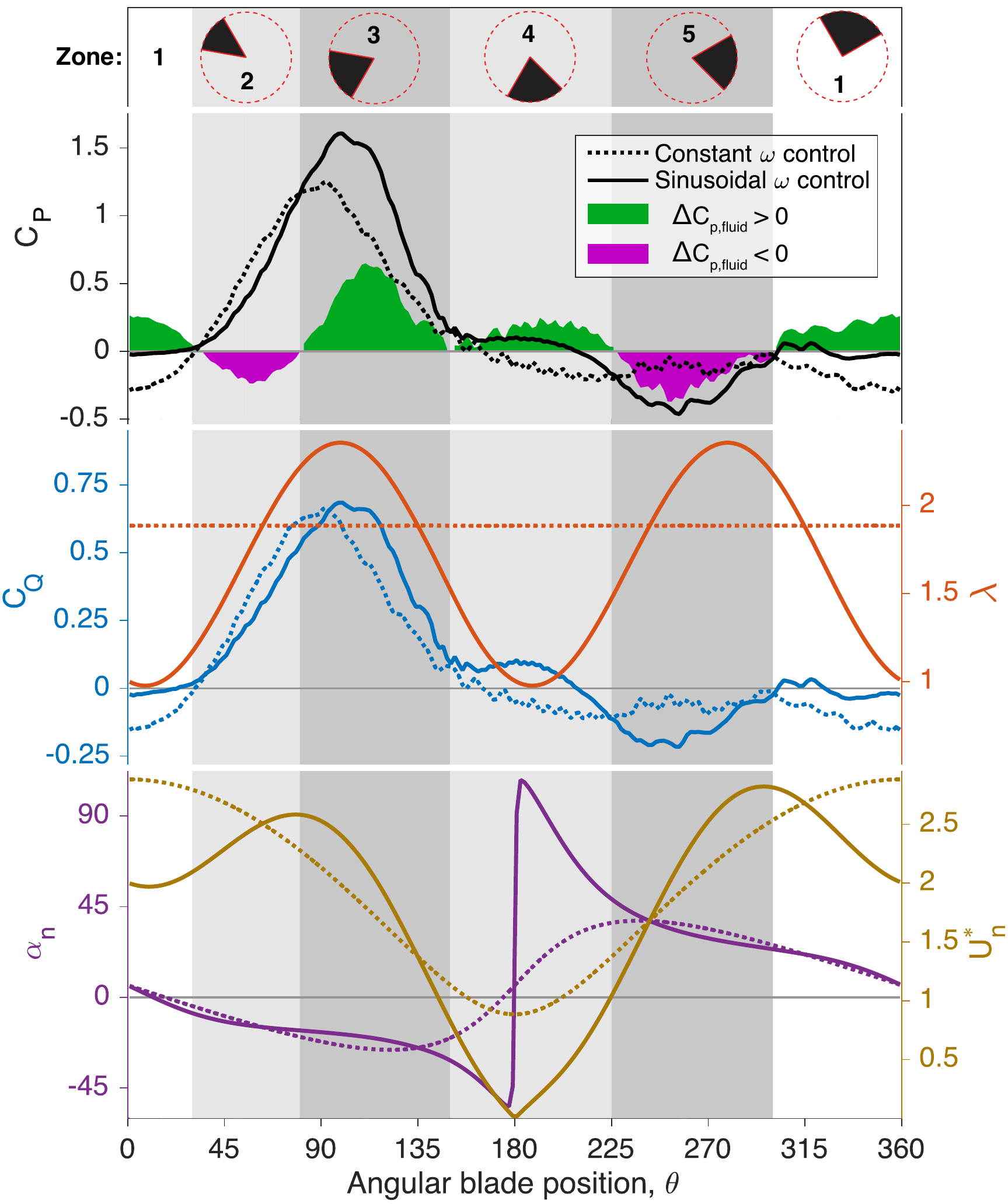}
\caption{\small \textsf{ Top: The phase averaged instantaneous efficiency profiles as a function of azimuthal blade position are compared for a one bladed turbine under constant and sinusoidal angular velocity control schemes given in Fig.~\ref{fig:results}. The instantaneous difference in efficiency is illustrated, with green and magenta areas indicating that the sinusoidal controller is performing better or worse than the constant velocity controller, respectively. Based on the differences in performance, the revolution of the blade is separated into five regions and the lower plots shows the operational context for these differences. The middle plot shows the torque coefficient, given by $C_Q = \tau \omega / (\frac{1}{2} \rho U_\infty^2 A R)$, in blue. The tip speed ratio profiles are given on the same plot in red. Because the free stream velocity is quasi-steady, this non-dimensionalizes the rotation rate.  The bottom plot shows the nominal angle of attack (purple) and the nominal free stream velocity (gold), which are calculated by Eqs.~\eqref{eq:alpha_n} and \eqref{eq:U_n}.  It should be noted, that because mean $C_P$ is found by integrating over time, rather than $\theta$, the sinusoidal control profiles are dilated or contracted, depending on the instantaneous rotation rate, but because this effect is small, it is more instructive to index performance to azimuthal blade position.}}\label{fig:zone}
\end{center}
\end{figure}

To investigate the mechanisms by which these increases are realized, results from a single-bladed turbine under constant and sinusoidal angular velocity control schemes identical to those listed in Fig.~\ref{fig:results} are compared. This isolates the fluid forcing to a single blade (see Sup. Fig. S6 for the validity of using a single bladed turbine as a proxy for the individual blade forcing of a two-bladed turbine). The semi-arbitrary control scheme cannot be explored with a single-bladed turbine due to structural instabilities that occur when large accelerations are applied to a turbine with an off-axis center of mass. Figure \ref{fig:zone} (top) shows the angular $C_P(\theta)$ profiles for the constant and sinusoidal angular velocity schemes implemented on a single-bladed turbine. Figure \ref{fig:zone} also shows the difference in efficiency between the sinusoidal and constant angular velocity control as a function of angular blade position. The net performance increase of the sinusoidal over the constant angular velocity control is examined as a function of azimuthal blade position. The blade stroke is broken into regions based on the variations in performance of the two controllers. 

	For $\theta > 180^\circ$, the blade passes through a region disturbed by its upstream passage. Consequently, $C_P$ is substantially reduced, even though the foil encounters favorable nominal angles of attack during both the upstream and downstream portions of the stroke.  Further, in this region, the assumptions underpinning these nominal values (e.g. inflow velocity comparable to the free stream velocity) are violated and the nominal values provide only a qualitative description of the hydrodynamics. Throughout, it should be recalled that the sinusoidal profile employed on this single-bladed turbine was optimized for the two-bladed turbine.  
	
	$\;\;$ \textbf{Zone 1, $\boldsymbol{\theta = 300^{\circ} \to 30^{\circ}}$} \hspace{0.25in} \\Here, a single-bladed turbine under constant angular velocity control is loosing energy (it has a negative efficiency), while the power produced by the sinusoidal angular velocity control scheme is near zero. The foil is translating almost directly up stream in this region.  This means that little of the lift produced is projected onto the direction of turbine rotation and, consequently, drag is likely to dominate the hydrodynamics. Because the turbine under sinusoidal velocity control is rotating slower than under constant velocity control, the nominal relative velocity is reduced, consequently reducing drag in the direction of rotation. 
	
	\textbf{Zone 2, $\boldsymbol{\theta = 30^{\circ} \to 80^{\circ}}$} \hspace{0.25in} \\During this portion of the rotation, the constant angular velocity control outperforms the sinusoidal controller.  Upon examining the instantaneous fluid torque coefficient curves (Fig.~\ref{fig:zone}, center, blue), it is apparent that the torque producing region is shifted to a later blade position under sinusoidal control. This shift may be due to a delay in the separation associated with the dynamic stall process. Possible causes include the following: under sinusoidal control, the foil is undergoing a virtual acceleration due to an increasing relative velocity, while the constant velocity controlled foil is decelerating (Fig.~\ref{fig:zone}, bottom, gold). In addition, the sinusoidal scheme has a smaller virtual pitch rate (Fig.~\ref{fig:zone}, bottom, purple).

	\textbf{Zone 3, $\boldsymbol{\theta = 80^{\circ} \to 150^{\circ}}$} \hspace{0.25in} \\ This region is associated with the majority of energy harvesting by the foils due to the favorable angle of attack and resulting dynamic stall forces. Here the sinusoidally controlled turbine enjoys the largest advantage. This is because power is the product of torque and angular velocity, and though the peak torques of the two schemes are similar, the angular velocity of the sinusoidal control scheme is at its maximum (as illustrated by the high tip speed ratio, $\lambda$, Fig.~\ref{fig:zone}, middle, red). This is striking because, even if the fluid forcing was identical between the two schemes, there would still be an increase in performance due the higher angular velocity under sinusoidal control. 
	
	\textbf{Zone 4, $\boldsymbol{\theta = 150^{\circ} \to 225^{\circ}}$} \hspace{0.25in}  \\Throughout this region, the sinusoidal scheme efficiency is usually positive, while the constant control scheme is consistently negative. Because the blade is translating directly downstream, the difference in performance cannot be attributed only to differences in the nominal angle of attack and the relative velocity.  Under sinusoidal control, the relative velocity is near-zero, limiting potential contribution of lift or drag from the free stream. However, this region does follow the main dynamic stall event in the rotation and the difference in blade forcing may be explained by improved interaction between the foil and the leading edge vortex. We note that the forcing due to these induced velocities are not considered in the nominal angle of attack approximations. 
	
	\textbf{Zone 5, $\boldsymbol{\theta = 225^{\circ} \to 300^{\circ}}$} \hspace{0.25in}  \\ The poorer performance of the sinusoidal control scheme during this region may be explained, in part, via the same mechanism as in zone 3. The slightly negative fluid torque, which is more negative than in the constant velocity scheme, is multiplied by a comparatively large angular velocity value, resulting in a more negative instantaneous efficiency. \\

	\section{Discussion}
	
	Though there is a complex interconnection between changes in the turbine kinematics, fluid forcing, and the resulting power output, the primary success of intracycle angular velocity control is derived from aligning maximum velocity with maximum torque generation. This may be analogous to other unsteady fluid control problems, such as a bird's perching maneuver. Secondary benefits are accrued by minimizing velocity when the turbine dynamics are drag dominated and exploiting interaction with the leading edge vortex generated during dynamic stall.
	
	Though the control scheme introduced here substantially increases turbine performance without actuators, control surfaces, or increasing the degrees of freedom, it is necessary to instantaneously supply the oscillatory power required to accelerate and decelerate the turbine. This has analogues to reactive power requirements for wave energy conversion~\cite{Falnes_2002}. Average power output may be smoothed by electrically connecting an array of turbines operating out of phase. Similarly, while a single turbine may require a larger generator for intracycle control, the mechanical coupling of multiple out-of-phase turbines may loosen this constraint.  
	
	Future work should explore the effectiveness of this control scheme on larger turbines operating at higher Reynolds number, as well as on turbines with alternative geometries. In addition, a number of refinements to the optimization process are possible. For example, additional terms in the definition of the angular velocity profile could allow the profile to approach a truly arbitrary waveform. In addition, variable angular velocity profile parameters may be optimized with objectives other than maximizing turbine efficiency. For instance, the optimization objective could seek a maximum $C_P$ constrained by peak thrust or lateral loads. Further, on-line optimization could be run continuously on a turbine, allowing it to adapt to slowly varying inflow conditions. This concept could be applied to optimization of arrays, to further enhance the type of bio-inspired array performance gains suggested by~\cite{Whittlesey_2010}.
	
	Historical experience with larger-scale cross-flow turbines show that efficiency with a constant velocity or torque controller is optimized for systems with a small number of blades (2-3) that have a small chord compared to the radius ~\cite{Weingarten_1976,Consul_2009}. These slender blades are prone to vibration and susceptible to damage from impact. In addition, these turbines operate at a higher tip speed ratio [ibid], which produces more noise, increases susceptibility to cavitation, and may pose an elevated risk of collision for avian and aquatic species. In this study, we have demonstrated that it is possible to achieve comparable efficiencies with a relatively high chord-to-radius ratio and relatively low rotational velocity. In parallel to refinements in turbine geometry, future optimization should explore the benefits of dynamic control to exploit unsteady fluid forces, as inspired by bio-propulsion, to realize transformative gains in efficiency for unconventional turbines.
		
	{\small
	\bibliographystyle{pnas2011}
	\bibliography{bibliog}

}



\section*{Supplementary material}

# Supplement

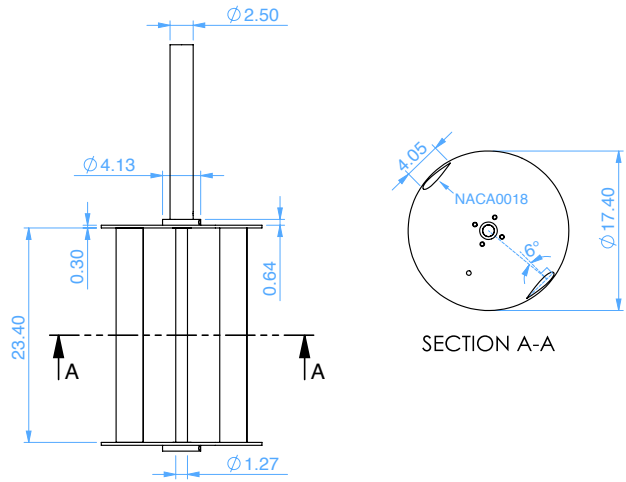

Figure S1: Detail diagram of the experimental turbine with measurement units in centimeters.

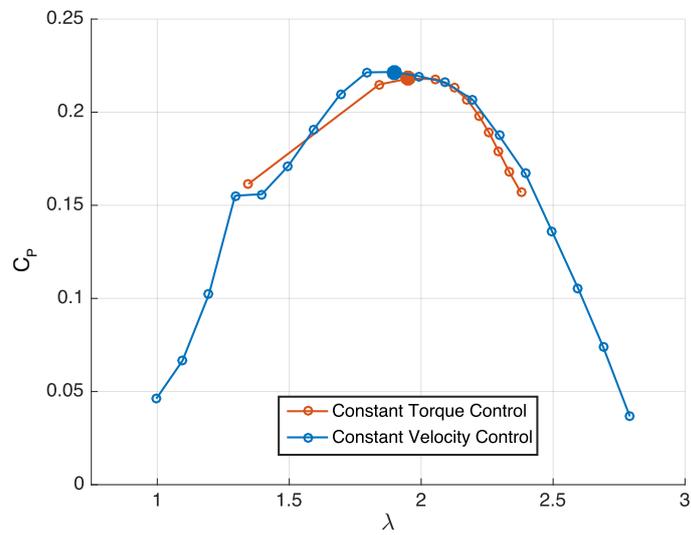

Figure S2: Performance curves of efficiency versus tip speed ratio for the two-bladed experimental turbine under constant angular velocity (blue) and constant resistive torque (red) control. The performance curves vary slightly due to the differing kinematics: in constant torque control the angular velocity oscillates slightly. In addition, the turbine is unable to operate at low tip speed ratios under torque control. The peak efficiency points (large dots) are used for comparison with intracycle angular velocity control.

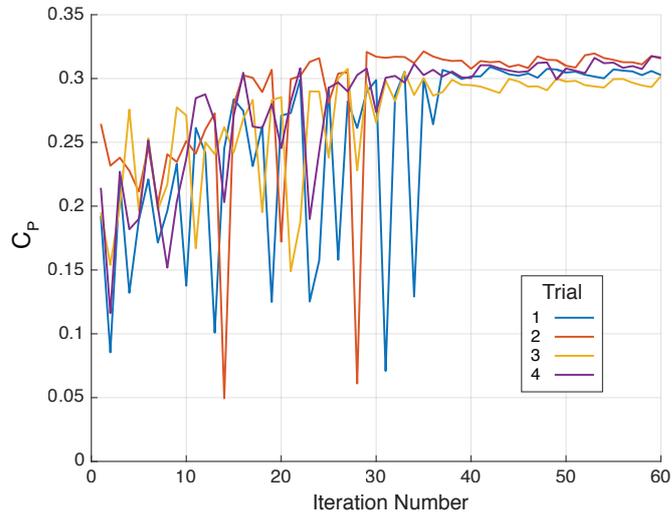
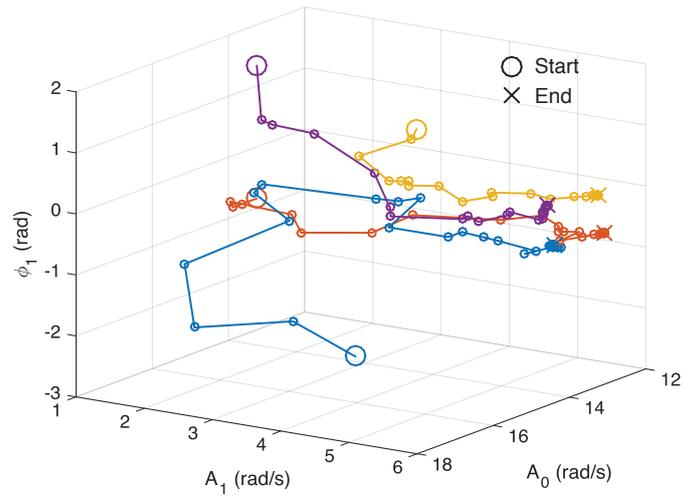

Figure S3: Top: Efficiency as a function of iteration number as the Nelder-Mead algorithm optimizes the three parameters that define the sinusoidal angular velocity control scheme. Bottom: The trajectories of the same optimization runs through the parameter space. Trajectories converge to a small region, but do not collapse to the same point. This may be a result of small variations in mean $C_P$ between function evaluations. The final optimal parameter set is taken as the mean ultimate parameter sets from individual optimization trials.

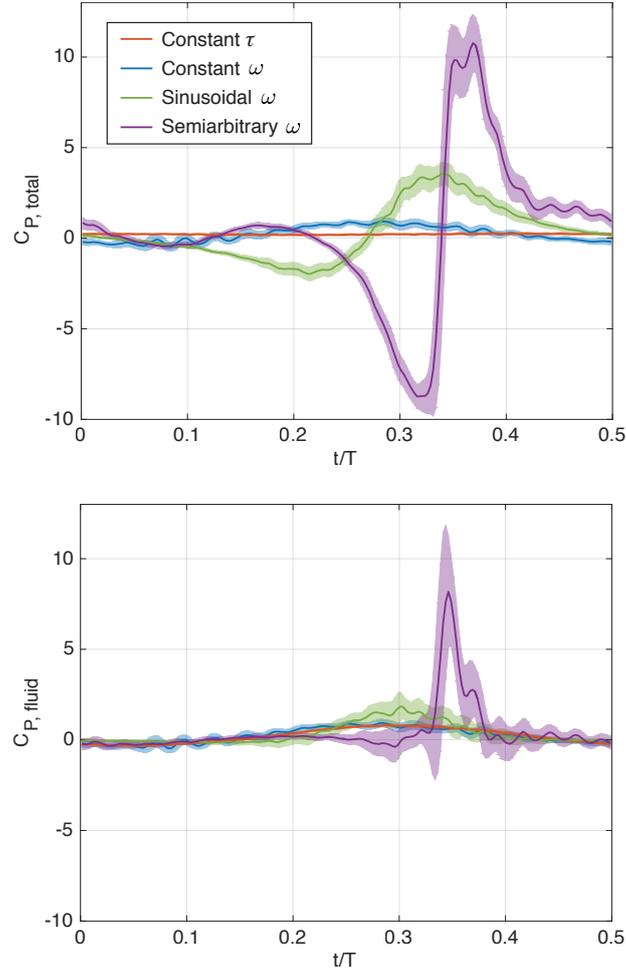

Figure S4: Here the instantaneous efficiencies (plotted with respect to time normalized by one rotation period, $T$) are shown with with (top) efficiency computed using the sum of the fluid torque and the torque associated with accelerating and decelerating the turbine and (bottom) using only the torque imparted by the fluid. The fluid efficiency, $C_{P,\text{fluid}} = C_P$ includes only the torque imparted on the turbine by the fluid and is used throughout this paper. The shaded regions indicate $\pm$ one standard deviation for the values that fall in the corresponding turbine position bin during the phase average calculation.

| Control Scheme | $\overline{C_P}_{,\text{total}}$ | $\overline{C_P}_{,\text{fluid}}$ | Difference |
| --- | --- | --- | --- |
| Constant $\tau$ | 0.218789 | 0.218931 | $1.4180 \times 10^{-4}$ |
| Constant $\omega$ | 0.229313 | 0.229324 | $1.0698 \times 10^{-5}$ |
| Sinusoidal $\omega$ | 0.321376 | 0.321384 | $8.1080 \times 10^{-6}$ |
| Semiarbitrary $\omega$ | 0.409662 | 0.409714 | $5.2025 \times 10^{-5}$ |

Figure S5: The mean efficiency (calculated over a many complete revolutions) is independent of the treatment of torque associated with the acceleration and deceleration of the turbine. This is because the rotational energy introduced to the turbine from the control forcing is periodic and conservative (because the angular velocity profiles are periodic). Thus, zero net energy is transferred to the turbine from the control system when multiple revolutions are considered.

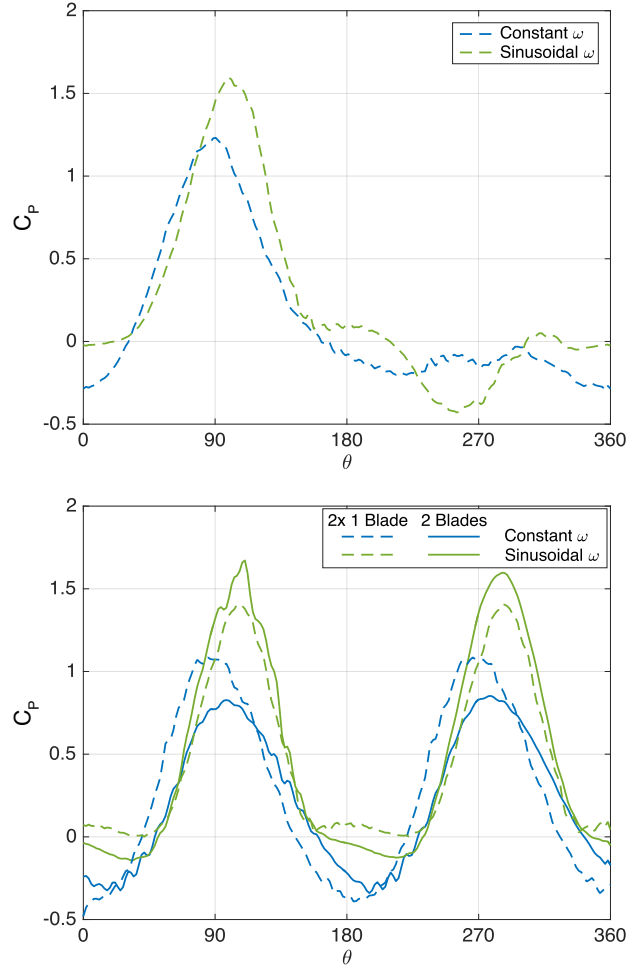

Figure S6: Two arguments are presented for the validity of examining the forcing on a single-bladed turbine as a proxy for the forcing on one blade of a two-blade turbine. The top plot shows the efficiency as a function of blade position for a single-bladed turbine under constant and sinusoidal velocity control. We note that the majority of power is produced during the first half of the cycle, despite the fact that during some regions of the downstream sweep of the blade ($\theta > 180°$) the nominal angle of attack is favorable. This is likely because the extraction of energy from the flow during the upstream portion of the blade stroke leaves little energy to be extracted on the downstream portion of the stroke. For a two-bladed turbine, with twice the solidity of a single-bladed turbine, this is also expected to occur. Therefore, the forces imparted to the blade during the upstream portion of the stroke, where the flow is relatively undisturbed by previous blade passes, should be relatively unaffected by the reduction of the number of blades from two to one. The lower plot shows a comparison between the efficiency profiles for the two-bladed turbine and a reconstruction of the efficiency profiles for a two-bladed turbine using single-bladed turbine data. The reconstruction adds the single-bladed efficiency to itself offset by half a rotation as follows: $C_{p,\ 2\text{x 1 Blade}} = C_{p,\ \text{one blade}}(0° \to 360°) + C_{p,\ \text{one blade}}(180° \to 360°, 0° \to 180°)$. As shown, good agreement is found between the reconstructed and actual efficiency profiles, with more phase error evident in the constant angular velocity control. In the future, individual blade instrumentation or flow field measurements near the blades may improve direct comparisons between control schemes.



\end{document}